\input{aipcheck}

\documentclass[
    ,final            
  ,numberedheadings
  ]
  {aipproc}
\usepackage{ulem} 
\usepackage{epsf}
\layoutstyle{6x9}

\def\beq{\begin{equation}}
\def\eeq{\end{equation}}
\def\bea{\begin{eqnarray}}
\def\eea{\end{eqnarray}}
\def\beqa{\begin{equation}\begin{array}{l}}
\def\eeqa{\end{array}\end{equation}}
\def\eqlab#1{\label{eq:#1}}
\def\figlab#1{\label{fig:#1}}

\def\eref#1{(\ref{eq:#1})}
\def\Eqref#1{Eq.~(\ref{eq:#1})}

\def\Figref#1{Fig.~\ref{fig:#1}}

\def\half{\mbox{\small{$\frac{1}{2}$}}}

\def\quarter{\mbox{\small{$\frac{1}{4}$}}}

\def\al{\alpha}
\def\be{\beta}
\def\ga{\gamma} 
\def\de{\delta} \def\De{\Delta}
\def\veps{\varepsilon}  \def\eps{\epsilon}

 \def\La{{\Lambda}}

\def\si{\sigma}

\def\dd{{\rm d}}
\def\pa{\partial}

\def\pa{\partial}

\def\nn{\nonumber}

\begin{document}

\title{Baryon  chiral perturbation theory---an update }

\classification{14.20.Dh, 14.20.Gk, 13.60.Fz, 13.60.Le, 11.30.Rd, 11.55.Fv}
 
\keywords      {light-by-light scattering, chiral expansion, dispersion relations}

\author{Vladimir Pascalutsa}{
  address={Institut f\"ur Kernphysik, Johannes Gutenberg Universit\"at, Mainz D-55099, Germany}
}

\begin{abstract}
The issue of consistent power counting in baryon chiral perturbation theory
is revisited.
\end{abstract}

\maketitle


\section{A prelude: low-energy light-by-light scattering}

The powerful effective-field theory (EFT) method allows for 
a perturbative treatment of the strong interaction at low 
energies  \cite{Weinberg:1978kz, Gasser:1983yg}. 
The perturbative expansion is made in powers of energy and momentum
of the lightest relevant degrees of freedom (pions, nucleons, etc.), rather than
in powers of the coupling constant as is usually done in (renormalizable) field theory. 

To illustrate the idea, imagine we would like to describe the properties
of cosmic microwave background (CMB), i.e., a gas of cold photons. 
The proper theory for this system is of course QED, and the interaction of
photons arises there through the electron loops. 
The typical energy of a CMB photon is 1 meV, which is much smaller  (by six
orders of magnitude!)  than the electron mass $m_e \approx 0.511$ MeV.
A spontaneous pair production is therefore excluded; the fermion degrees
of freedom are too heavy and will never appear explicitly in this system.
It is possible to eliminate them from the description altogether by 
integrating out the fermion fields and expanding the result in powers of inverse electron
mass. Another way of doing that is to write an ``effective theory''
containing only the relevant degrees of freedom, {\it viz.}, the photon field $A_\mu(x)$. 
The most general Lagrangian of such theory would begin with
\beq
\eqlab{EH}
\mathcal{L} = -\quarter F_{\mu\nu}F^{\mu\nu} + a_1 \pa_\al F_{\mu\nu} \pa^\al F^{\mu\nu}
+ c_1 ( F_{\mu\nu}F^{\mu\nu})^2 + c_2 ( F_{\mu\nu}\tilde F^{\mu\nu})^2 +\ldots, 
\eeq 
where $F$ is the field strength and $\tilde F$ is its dual. 
The first term in this Lagrangian is just the free electromagnetic (e.m.) radiation
and the rest represents the effect of electron loops at low energies. 
Even if this effect is not known to us precisely, we know that it will satisfy all the symmetries of QED, such as the Lorentz- and discrete symmetries, as well as the electromagnetic gauge invariance. 
These symmetries allow the effective theory to be written in terms
of $F_{\mu\nu}$ only, and as the result, the Lagrangian can be ordered in powers 
of derivatives of the photon field. A derivative translates into the momentum, or energy, and
hence the derivative expansion translates into the expansion in energy.
The parameters of the effective theory, in this case $a$'s and $c$'s, must be expressed
in terms of the QED parameters---electron mass and charge. This can be achieved by
``matching": calculating the same quantity in the effective theory and in QED, and equating the
results. For example, $a_1$ can be determined from the vacuum polarization, while
$c_1$ and $c_2$ can be matched at the level of light-by-light scattering 
amplitudes.

The effective framework is especially useful when the underlying theory is non-perturbative,
as it is in the case of QCD at low energies. In our example, 
the hadronic effects in the photon gas can still be presented in the form of  \Eqref{EH}.
It is still hard to calculate from first principles what the hadronic contribution
to the low-energy constants is, but we can measure it! 

For example, the constants $c_1$ and $c_2$, describing the low-energy 
photon self-interactions, can  be related to linearly-polarized cross-sections of photon-photon 
fusion,  $\si_{||}$ and $\si_{\perp}$ \cite{Pascalutsa:2010sj}:
\beq
c_1 = 
 \frac{1}{8\pi }\int_{0}^{\infty}\! \dd s\,  \frac{ \si_{||} (s) }{s^2}, \qquad
 c_2 
 = 
 \frac{1}{8\pi }\int_{0}^{\infty} \!\dd s\,  \frac{  \si_\perp(s)}{s^2},
\eeq
where $\sqrt{s}$ is the total invariant energy of the $\ga\ga$ collision.
From this we know already that these constants are positive definite and so
the low-energy photons attract. 
Taking the cross-sections of $\ga\ga\to\, $hadrons, we can obtain
the hadronic contribution to these quantities, and with that we can calculate 
the most important hadronic effects in the photon gas. 
I have to note though that the polarized  $\ga\ga$ fusion cross-sections
have not yet been measured and so this example for now is \sout{academic}  served
for illustration purpose only.  

\section{Versions of Baryon $\chi$PT and the nucleon mass}
 
 If we replace photons with pions in the above example, and electrons with quarks and gluons,
 we must end up with the chiral EFT,  commonly referred to as chiral perturbation
 theory ($\chi$PT). The name $\chi$PT was originally assigned to the expansion of
 static quantities, such as masses, electromagnetic moments, in powers of the pion mass
 \cite{Pagels:1974se}. In modern language,  $\chi$PT is a simultaneous expansion in
 powers of pion mass and momentum---the chiral expansion. 
 The break-down scale of this expansion is
 believed to be set at around 1 GeV. 
 
 Introducing the nucleon into the picture, in the words of first paper attempting it
 \cite{GSS89}, ``complicates life a lot." First of all, the nucleon is heavy, and its
 mass seems to pop out in the places it shouldn't, violating some power-counting
 arguments; we will see one example in a moment.
 Secondly, the nucleon is easily excited into the $\De(1232)$, the excitation energy
 being $\Delta \equiv M_\De-M_N \simeq 293$ MeV. 
 In attempts to find a systematic treatment of these issues, several
 different versions of baryon $\chi$PT were born:
 \begin{itemize}
 \item Heavy-baryon $\chi$PT (HB$\chi$PT)  \cite{JeM91a}, where in addition to the
 chiral expansion, a semi-relativistic expansion in the inverse nucleon mass is made.
 \item Infrared Regularization (IR-B$\chi$PT) \cite{Becher:1999he}, where the negative-energy
 pole is removed from the baryon propagators.
 \item Extended on-mass-shell scheme (EOMS-B$\chi$PT) \cite{Fuchs:2003qc}, which
 recognizes that certain renormalizations must be done to implement 
 consistent power counting.
 \end{itemize}
Let us see how these versions work on
 the elementary example of the nucleon mass.

 The chiral expansion of the nucleon mass to order $p^3$ is given by
 \beq
 \eqlab{nmass}
 M_N = \stackrel{\circ}{M_N} - 4 \stackrel{\circ}{c_1} m_\pi^2 + \Sigma^{(3)}_N, 
 \eeq
 where $\stackrel{\circ}{M_N}$ and $\stackrel{\circ}{c_1}$ are low-energy constants
 (LECs) from the chiral effective Lagrangian, 
 and $\Sigma$ is the nucleon self-energy. The nucleon self-energy may have (infinitely) many
 chiral-loop contributions of the type shown in \Figref{nuclse}, however the Weinberg's
  power counting 
 \begin{figure}[tb]
 \centerline{\epsfclipon   \epsfxsize=12.5cm%
  \epsffile{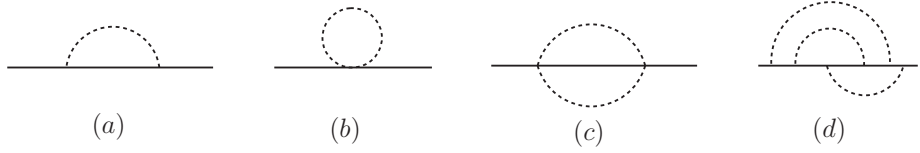} 
}
\caption{Graphs representing the chiral-loop corrections to the nucleon mass. Nucleon (pion) propagators are denoted by solid (dashed) lines.}
\figlab{nuclse}
\end{figure}
tells us that a graph with 
$L$ loops, $N_\pi$ pion and $N_N$ nucleon lines, 
$V_k$ vertices from the Lagrangian of order $k$, 
 contributes to order $p^n$, with 
 \beq
\eqlab{pc}
n =4L - 2N_\pi -N_N +  \sum\nolimits_k  k V_k \, .
\eeq
The leading $\pi N$ couplings (pseudovector, Weinberg--Tomozawa, etc.) 
are of the first order ($k=1$)
and therefore, to order $p^3$ only the graph (a) contributes.
Evaluating this graph yields~\cite{Ledwig:2010nm}:
\bea
 \Sigma^{(3)}_N  & =&  \frac{3 g_A^2 }{(4\pi f_\pi)^2}\Bigg\{ - \half L_\veps M_{N}^3
+\half \left(1-L_\veps\right) M_N\, m_\pi^2 
\nn \\
&& 
 \, - \,\,  m_\pi^{3}\,  \sqrt{1-\frac{m_\pi^2}{4M_N^2} } \, \,\arccos \frac{m_\pi}{2M_N}  -
 \frac{m_\pi^4}{2M_N}\,  \ln \frac{m_\pi}{M_N} 
  \Bigg\}.
  \eqlab{renorm} 
\eea
where  $L_\veps = -1/\veps -1 + \gamma_E - \ln(4\pi \La/M_{N})$
exhibits the UV divergence as $\veps=(4-d)/2 \to 0$, with $d$ being the number of space-time dimensions, $\La$  the scale of dimensional regularization, and $\gamma_{E}\simeq 0.5772$ the Euler's constant. 
For simplicity we have assumed the physical values for the parameters entering the loop: 
$M_N\simeq 939$ MeV, $g_A\simeq 1.267$, $f_\pi\simeq 92.4$ MeV;
the difference with the chiral-limit values leads to  higher-order effects.

The expression \eref{renorm} is an exact result of a textbook calculation and so it is
disappointing to see that it seems to invalidate the power counting formula \eref{pc}. The power counting estimates this loop is of $p^3$ size, in this case $m_\pi^3$, but the first
two terms are obviously larger. On the other hand, this expression begs for a renormalization.
We do have the two low-energy constants in \Eqref{nmass}, which can be renormalized to
absorb the UV divergencies and remove the dependence of the nucleon mass on the renormalization scale. However, only having $L_\eps =0$ in
\Eqref{renorm}, which corresponds to the $\overline{\mbox{MS}}$ scheme
applied in the original paper \cite{GSS89}, does not work:
there is the $M_N m_\pi^2$ term remaining, which violates the power counting.
The essence of the EOMS \cite{Fuchs:2003qc} is to absorb this  term
too in the course of the renormalization. As the result we have,
\bea
&&  \mbox{EOMS}: 
 \, M_N =\,\, \stackrel{\circ}{M_N} -\, 4 \stackrel{\circ}{c_1} m_\pi^2 
-  \frac{3 g_A^2 m_\pi^3 }{(4\pi f_\pi)^2}\Bigg\{
  \sqrt{1-\mbox{$\frac{m_\pi^2}{4M_N^2}$} } \, \,\arccos 
  \mbox{$\frac{m_\pi}{2M_N}$}  +
 \frac{m_\pi}{2M_N}\,  \ln \frac{m_\pi}{M_N} 
  \Bigg\}, \nn\\
  &&  \mbox{IR}:
\,   M_N =\,\, \stackrel{\circ}{M_N} -\, 4 \stackrel{\circ}{c_1} m_\pi^2 
 - \frac{3 g_A^2 m_\pi^3 }{(4\pi f_\pi)^2}\Bigg\{
  \sqrt{1-\mbox{$\frac{m_\pi^2}{4M_N^2}$} } \, \,\arccos 
 \Big(-  \mbox{$\frac{m_\pi}{2M_N}$} \Big) +
 \frac{m_\pi}{2M_N}\,  \ln \frac{m_\pi}{M_N} \nn\\
 && \qquad \qquad +\,  (L_\eps-1) \frac{m_\pi}{4M_N}
  \Bigg\},
   \eqlab{schemes} \\
&&  \mbox{HB}: M_N =\,\, \stackrel{\circ}{M_N} -\, 4 \stackrel{\circ}{c_1} m_\pi^2 
 - \frac{3 g_A^2 m_\pi^3 }{(4\pi f_\pi)^2}\, \frac{\pi}{2}\,, \nn
\eea
where $\stackrel{\circ}{M_N}$ and $\stackrel{\circ}{c_1}$
are now the renormalized (physical) values of these parameters, and where, for comparison, I displayed also the $p^3$ result of the IR and HB versions of
B$\chi$PT.

The EOMS result is consistent with power counting and the $\overline{\mbox{MS}}$
 is not, but
can one renormalization scheme be better than the other? In EFT it can.
Observe that the whole
difference between the two schemes at this order is expressed
as the difference in the value of $c_1$:
\beq
\stackrel{\circ}{c_1}^{\!(\mathrm{\overline{MS}} )}\!\! - \,\, \frac{3 g_A^2 M_N}{8 (4\pi f_\pi)^2} \,=\quad  \stackrel{\circ}{c_1}^{\!(\mathrm{EOMS} )}
\eqlab{diff}
\eeq
Suppose now we match $\stackrel{\circ}{c_1}$ to QCD by fitting, for instance, to 
the lattice QCD data. In the two schemes the fit will be identical but the
values for the LEC will differ according to \Eqref{diff}. It is still possible that 
both $c_1$ are of  {\it natural} size, i.e., of oder of one in GeV units. Now,
after calculating to one order higher, we would refit the data and find a new
value of the LEC. In EOMS the value of $c_1$ would not change much
from one order to another, since $M_N$ as a function of $m_\pi$ would indeed
change only in the higher-order terms. In the $\overline{\mbox{MS}}$ scheme 
the value of $c_1$ could change a lot, since any loop 
can produce a large shift in that $m_\pi^2$ term. The latter situation is 
not satisfactory, especially if we want to have our LECs to represent some
physical quantities. For example, we would like $\stackrel{\circ}{M_N}$
to be the nucleon mass in the chiral limit, while $\stackrel{\circ}{c_1}$ could
represent the value of the $\si$-term in the chiral limit. In this case the EOMS
is indeed favored over the $\overline{\mbox{MS}}$. 

Looking back at the other results displayed in \Eqref{schemes}, 
we ought to dismiss the IR scheme for not having the correct analytic properties.
The non-analyticity of the square root at $m_\pi = 2M_N$ is canceled
in the EOMS due to the factor of $\arccos(1) =0$, and is not canceled
in the IR, because  $\arccos(-1) =\pi$. Of course, at small pion masses
this pathology is barely seen, as  the IR result is different from 
the EOMS by the following term alone: 
\beq
\de M_N =  - \frac{3 g_A^2  }{(4\pi f_\pi)^2}\frac{m_\pi^4}{M_N}
\Big\{  \sqrt{1-\mbox{$\frac{m_\pi^2}{4M_N^2}$} } 
+ \quarter (L_\eps-1)
  \Big\}.
\eeq
One can argue that this 
term is of order $p^4$, which is beyond the accuracy of this calculations.
But then the question is why do we need this term at all, especially when
it occurs as a result of violation of the analytic properties.

We finally come to the point that the EOMS expression in \Eqref{schemes}
contains 
an infinite amount of terms which are nominally of higher order in $p\sim m_\pi$,
while in the HB expression  these terms are happily dropped. I deliberately
used the word ``nominally", because a term going as $m_\pi^4$ is not
necessarily smaller then a $m_\pi^3$ term---depends on the coefficients.
One assumes the coefficients to be all of natural size, but is this always true?

We can see this to be pretty much true for our case. Expanding the 
 factor in curly brackets in the EOMS expression of \Eqref{schemes}, we find
 \beq
 \Big\{ \ldots\Big\}  = \frac{\pi}{2} + \frac{m_\pi}{2M_N} \Big( \ln \frac{m_\pi}{M_N} -1 \Big)
 -\frac{\pi m_\pi^2}{16M_N^2} + O(m_\pi^3), 
 \eeq
 and hence the HB result, $\pi /2$, is a very good {\it approximation}.
 But then there are other examples. 
 The magnetic polarizability of the proton $\be_p$ at order $p^3$ 
 in B$\chi$PT expands as \cite{Bernard:1991rq,Lensky:2009uv}:
 \beq
 \beta_p = \frac{\al_{em} g_A^2}{192 \pi f_\pi^2 M_N}
 \left\{ \frac{M_N}{m_\pi} + \frac{72}{\pi} \ln \frac{m_\pi}{M_N} + \frac{126}{\pi}
  - \frac{981}{8}\frac{ m_\pi}{M_N} + O(m_\pi^2) \right\}.
 \eeq
 One can see that the nominally-higher-order terms have unnaturally
 large coefficients and cannot be neglected. 
 The same situation is observed in the other polarizabilities of the nucleon. 
 In such cases the HB expansion fails. 
 
A comprehensive comparison of the various schemes has recently been
done by Geng et al.~\cite{Geng:2009hh,Geng:2009ys} 
in the context of the SU(3) B$\chi$PT study of the baryon magnetic moments.
The EOMS comes out to be favored by this study as well.
As for the other recent applications of B$\chi$PT in the on-mass-shell scheme, 
I would like to mention a next-to-next-to-leading order calculation of the proton Compton scattering
\cite{Lensky:2009uv},
and of the nucleon and $\De(1232)$-isobar electromagnetic form factors
\cite{Ledwig:2011iw}.





\begin{thebibliography}{9}
\bibitem{Weinberg:1978kz}
  S.~Weinberg,
  Physica A {\bf 96}, 327 (1979).

\bibitem{Gasser:1983yg}
  J.~Gasser and H.~Leutwyler,
  Annals Phys.\  {\bf 158} (1984) 142.
  
\bibitem{Pascalutsa:2010sj}
  V.~Pascalutsa and M.~Vanderhaeghen,
  Phys.\ Rev.\ Lett.\  {\bf 105}, 201603 (2010)
  [arXiv:1008.1088].
  
\bibitem{Pagels:1974se}
  H.~Pagels,
  Phys.\ Rept.\  {\bf 16}, 219 (1975).
  
  \bibitem{GSS89}
J.~Gasser, M.~E.~Sainio and A.~Svarc,
Nucl.\ Phys.\ B {\bf 307}, 779 (1988).

    \bibitem{JeM91a}
E.~Jenkins and A.~V.~Manohar,
Phys.\ Lett.\ B {\bf 255}, 558 (1991).

\bibitem{Becher:1999he}
  T.~Becher and H.~Leutwyler,
  Eur.\ Phys.\ J.\  C {\bf 9}, 643 (1999)
  [arXiv:hep-ph/9901384].
  
   \bibitem{Fuchs:2003qc}
  T.~Fuchs, J.~Gegelia, G.~Japaridze and S.~Scherer,
  Phys.\ Rev.\  D {\bf 68}, 056005 (2003).
  
\bibitem{Ledwig:2010nm}
  T.~Ledwig, V.~Pascalutsa and M.~Vanderhaeghen,
  Phys.\ Lett.\  B {\bf 690}, 129 (2010).
  
  \bibitem{Bernard:1991rq}
  V.~Bernard, N.~Kaiser and U.-G.~Mei{\ss}ner,
  Phys.\ Rev.\ Lett.\  {\bf 67}, 1515 (1991).
  
\bibitem{Lensky:2009uv}
  V.~Lensky, V.~Pascalutsa,
  Eur.\ Phys.\ J.\  C {\bf 65}, 195 (2010).

\bibitem{Geng:2009hh}
  L.~S.~Geng, J.~Martin Camalich, M.~J.~Vicente Vacas,
  Phys.\ Lett.\  {\bf B676}, 63-68 (2009).
 
\bibitem{Geng:2009ys}
  L.~S.~Geng, J.~Martin Camalich, M.~J.~Vicente Vacas,
  Phys.\ Rev.\  {\bf D80}, 034027 (2009).

\bibitem{Ledwig:2011iw}
  T.~Ledwig, J.~Martin-Camalich, V.~Pascalutsa, M.~Vanderhaeghen,
  arXiv:1105.0468; in preparation.
\end{thebibliography}
\end{document}